\documentclass[a4paper,12pt]{article}
\usepackage{amsmath}
\usepackage{amsfonts}
\usepackage{amssymb}
\usepackage{marvosym}
\usepackage{wasysym}
\usepackage{caption}
\usepackage{subcaption}
\usepackage{fullpage}
\usepackage[]{graphicx}
\usepackage{hyperref}
\usepackage[english]{babel}
\usepackage{xcolor}
\usepackage{makeidx}

\hypersetup{citebordercolor=0 1 0, linkbordercolor=1 1 1,}
\begin{document}

\begin{titlepage}
\title{Conversion of relic gravitational waves into photons in cosmological
magnetic fields}
\author{Alexander D. Dolgov$^{1,2,3,4}$ and Damian Ejlli$^{1,2,5}$}
\date{}
\maketitle
\begin{center}
$^{1}$\emph{Dipartimento di Fisica e Scienze della Terre, Universit\'{a} degli Studi di Ferrara, \\
Polo Scientifico e Tecnologico-Edificio C, Via Saragat 1, 44122 Ferrara, Italy}\\
$^{2}$ \emph{Istituto Nazionale di Fisica Nucleare (INFN), Sezione di Ferrara, \\
Polo Scientifico e Tecnologico-Edificio C, Via Saragat 1, 44122 Ferrara, Italy}\\
$^{3}$\emph{Department of Physics,Novosibirsk State University, \\Pirogova 2, Novosibirsk 630090, Russia\\
$^{4}${ITEP}, Bol. Cheremushkinskaya 25, Moscow 117218 Russia} \\
$^{5}$\emph{Astroparticule et Cosmologie (APC),
Universit\'{e} de Paris Diderot-Paris 7\\
10, rue Alice Domon et L\'{e}onie Duquet,
75205 Paris Cedex 13
France}
\end{center}
\thispagestyle{empty}

\begin{abstract}
Conversion of gravitational waves into electromagnetic radiation is discussed. The probability of transformations of gravitons 
into photons in presence of cosmological background magnetic field is calculated at the recombination epoch and during  
subsequent cosmological stages. The produced electromagnetic radiation is concentrated in the X-ray part of the spectrum. 
It is shown that if the early Universe was dominated by primordial black holes (PBHs) prior to Big Bang Nucleosynthesis (BBN), 
the relic gravitons emitted by PBHs would transform to an almost isotropic background of electromagnetic radiation due to 
conversion of gravitons into photons in cosmological magnetic fields. Such extragalactic radiation could be noticeable
or even dominant component of Cosmic X-ray Background.
\end{abstract}

\vspace{5cm}
\Email{Alexander D. Dolgov:  \href{mailto:dolgov@fe.infn.it}{\nolinkurl{dolgov@fe.infn.it}} }\\

\Email{Damian Ejlli: \href{mailto:ejlli@fe.infn.it}{\nolinkurl{ejlli@fe.infn.it}}}

\end{titlepage}

\pagenumbering{arabic}

\section{Introduction}\label{sec:1}

During last decades, many space detectors have been exploring the Universe in different energy bands revealing rich flavors of 
radiation coming from different regions of the Universe.  These multi messengers not only tell about their region of emission, 
but also give unique opportunity to test the laws of physics and probe their limits. One of the most important multi messengers 
that has been detected is, without any doubt, the Cosmic Microwave Background Radiation (CMBR) which is an unique probe 
of the Universe at the last scattering surface and, moreover, it tells about much earlier stages of the cosmological evolution.
Other messengers from the early and the present-day Universe include cosmological neutrino background, axions, magnetic 
monopoles, cosmic rays, extragalactic $\gamma$-rays, cosmic $X$-rays, and gravitational waves. In this paper we study  
connection between electromagnetic radiation (photons) and gravitational waves (gravitons) produced in the early universe.

The search for gravitational radiation is one of the central problems of General Relativity and modern cosmology.
Predicted by Einstein in 1918~\cite{Einstein:1918cc}, gravitational waves (GW) still escape observations,  
though an indirect evidence of gravitational radiation was discovered by a decrease of the orbital period of the double
pulsar PSR B1913+16~\cite{hulse-taylor}. At the present time there is an active search for gravitational radiation from 
many catastrophic astrophysical phenomena as well as for the cosmological background of relic gravitational waves,
for a recent review see e.g. ref.~\cite{gw-rev}.

In this paper we concentrate on relic high frequency GWs of cosmological origin.
A study of cosmological GWs was initiated in ref.~\cite{grischuk}, where graviton production by time-dependent
curvature was considered. In ref.~\cite{star79} it was found that gravitational waves are efficiently produced during
inflation (see also~\cite{rubakov-gw}). The frequency of such waves in the interesting for detection range (and also 
of  those created by many other cosmological sources) is quite low, i.e. a fraction of Hz.
The existing ground based interferometers (VIRGO, LIGO etc) and space detectors (LISA, DECIGO, BBO etc) are
most sensitive in this low frequency part of the spectrum. 

In refs.~\cite{ Dolgov:2011cq} a different mechanism of the relic GW generation in the early Universe was studied, which could produce gravitational waves in GHz and much higher frequency range and quite possibly with high 
density parameter $\Omega_{\textrm{gw}}$.  Such GWs would be produced if at some early epoch the universe was 
dominated by primordial black holes (PBH) and GWs originated from their evaporation and/or from coalescence of 
the PBH binaries. 

The frequency of such GWs is by far beyond the sensitivity range of the present traditional detectors but 
they are possibly detectable by future high frequency electromagnetic gravitational detectors . Moreover, their 
registration does not look very difficult due to the process of GW transformation into photons in 
external magnetic field.  In pioneering paper~\cite{gertsen} the inverse process of photon to graviton transformation
in magnetic field was studied. It followed by several works dedicated to graviton to photon transition~\cite{g-to-gamma}.
Below we calculate the probability of GWs transformation into electromagnetic radiation in the primordial magnetic field
after the hydrogen recombination and argue that the registration of this radiation might be feasible.
The presented below  calculations of the intensity of the electromagnetic radiation
closely follow those described in ref.~\cite{Raffelt:1987im}.

The paper is organized as follows. In sec.  \ref{sec:2} we derive equations of motions for the graviton-photon system in 
static magnetic field in the case when the graviton wavelength is smaller than magnetic field coherence length. 
In sec. \ref{sec:3} we derive the mixing probability for graviton to photon oscillation using the wave function approximation. 
The oscillation probability is analogous to oscillation between neutrinos with different flavors. 
In sec. \ref{sec:4}   we calculate oscillation probability at a qualitative level using the wave function approximation
at various stages during the cosmological history, 
namely at recombination and the contemporary epoch. 
In sec. \autoref{sec:5} we do the same thing as in sec. \ref{sec:4}, using the 
 density matrix approach in order to take into account coherence breaking due to scattering of photons 
in  plasma. In sec. \ref{sec:6} we discuss observable effects of graviton to 
photon conversion at the present time and in sec. \ref{sec:7} we conclude.

\section{Equations of motions of graviton-photon system}\label{sec:2}

The total action describing gravitational and electromagnetic fields is given by
the sum of two terms:
\begin{equation}\label{totalaction}
 \mathcal S=\mathcal S_g+\mathcal S_{em},
\end{equation}
where $\mathcal S_g$ is the usual Einstein-Hilbert action equal to:
\begin{equation}\label{gravaction}
\mathcal S_g=\frac{1}{\kappa^2}\int \mathrm d^4x\sqrt{-g} R
\end{equation}
and $\mathcal S_{em}$ is the action of the electromagnetic field minimally coupled to gravity:
\begin{equation}\label{mattaction}
\mathcal S_m =-\frac{1}{4}\int \mathrm d^4x\sqrt{-g}\,g^{\mu\rho}g^{\nu\sigma}F_{\mu\nu}F_{\rho\sigma} + \frac{\alpha^2}{90m_e^4}\int\mathrm d^4x\sqrt{-g}\,[(F_{\mu\nu}F^{\mu\nu})^2+\frac{7}{4}(\tilde{F}_{\mu\nu}F^{\mu\nu})^2]\,.
\end{equation}
The first term above is the Maxwell action and the second quartic one is the Heisenberg-Euler contribution~\cite{EH-Schwinger} originating
from the electron box diagram. 
This term  describes nonlinear corrections to the classical electrodynamics in the limit of low
photon frequencies, $\omega\ll m_e$. As we will see below the second term gives the photon an effective refraction index 
in vacuum with external magnetic field.  Using this expression for the action confines the validity of the results presented 
here to sufficiently low frequencies. However, they can be easily generalized to higher $\omega$.

Here we use the natural units, $c=\hbar=k_B=1$. The essential quantities are defined as follows:
 $\kappa\equiv\sqrt{16\pi G}$, $G\equiv 1/m_{Pl}^2$ is the Newton constant, 
 $m_{Pl} = 1.2\cdot 10^{19}$ GeV\footnote{{In the literature another notation is often used, namely, $\kappa =8\pi G$.}} ,
$\alpha=e^2/(4\pi)$ is the fine structure constant, $m_e$ is the electron mass, $g_{\mu\nu}$ is the metric tensor with signature 
$g_{\mu\nu}=\textrm{diag}(-1, 1, 1, 1)$, $g=\textrm{det}(g_{\mu\nu})$, 
%is the metric determinant, 
$R$ is the curvature scalar defined as $R=g^{\mu\nu}R_{\mu\nu}$ with $R_{\mu\nu}$ being
the Ricci tensor, $R_{\mu\nu} = R^\alpha_{\mu\alpha\nu}$;
the Riemann tensor in this paper is defined as in Misner, Thorne and Wheeler~\cite{Misner:1974qy}, 
$ R^{\mu}_{\;\nu\rho\sigma}=\partial_{\rho}\Gamma_{\nu\sigma}^{\mu}-\partial_{\sigma}\Gamma_{\nu\rho}^{\mu}+
\Gamma_{\alpha\rho}^{\mu}\Gamma_{\nu\sigma}^{\alpha}-\Gamma_{\alpha\sigma}^{\mu}\Gamma_{\nu\rho}^{\alpha} $,
where the Christoffel symbol is $\Gamma_{\mu\nu}^{\alpha}=
\frac{1}{2}g^{\alpha\sigma}(\partial_\mu g_{\sigma\nu}+\partial_\nu g_{\sigma\mu}-\partial_\sigma g_{\mu\nu}) $. 
$F_{\mu\nu}$ is the electromagnetic field strength tensor and 
$\tilde{F}_{\mu\nu}\equiv ({1}/{2})\epsilon_{\mu\nu\rho\sigma}F^{\rho\sigma}$ is its dual. 
The metric tensor of a weak gravitational wave propagating in flat space-time can be written as follows:
\begin{equation}\label{metricsplit}
 g_{\mu\nu}=\eta_{\mu\nu}+\kappa h_{\mu\nu}(\mathbf{x}, t),
\end{equation}
where $\eta_{\mu\nu}$ is the flat Minkowski metric tensor and $h_{\mu\nu}$ are small perturbation around flat space-time, $|h_{\mu\nu}|\ll 1$. 
Considering terms up to the second order in $h_{\mu\nu}$ 
%the coupling constant $\kappa$, 
we rewrite gravitational action \eqref{gravaction} 
as
\begin{equation}\label{lingravact}
\mathcal S_g=-\frac{1}{4}\int \mathrm d^4x\,[\partial_\mu h_{\alpha\beta}\partial^{\mu}h^{\alpha\beta}-\partial_\mu h\partial^\mu h+2\partial_\mu h^{\mu\nu}\partial_\nu h-2\partial_\mu h^{\mu\nu}\partial_\rho h_\nu^{\rho}]
\end{equation}
and electromagnetic part \eqref{mattaction} becomes:
\begin{equation}\label{linmattact}
\mathcal S_{em} =-\frac{1}{4}\int \mathrm d^4x\,F_{\mu\nu}F^{\mu\nu}+
\frac{\kappa}{2}\int \mathrm d^4x\, h_{\mu\nu}T^{\mu\nu}+ 
\frac{\alpha^2}{90m_e^4}\int\mathrm d^4x\,[(F_{\mu\nu}F^{\mu\nu})^2+\frac{7}{4}(\tilde{F}_{\mu\nu}F^{\mu\nu})^2],
\end{equation}
where $T_{\mu\nu}$ is the electromagnetic energy-momentum tensor, 
$T^{\textrm{em}}_{\mu\nu}=F_{\mu\rho}F^\rho_\nu-\eta_{\mu\nu}F_{\alpha\beta}F^{\alpha\beta}/4$. 
Total Lagrangian density  \eqref{totalaction} is given by the sum of  linearized actions \eqref{lingravact} and \eqref{linmattact}.

The Euler-Lagrange equation of motion for fields $h_{\mu\nu}$ and $A_\mu$ are obtained  
by taking the variation of the total action with respect to these fields with usually
imposed the Traceless Transverse (TT) gauge condition: 
$h_{0\mu}=0, \partial_j h_{ij}=0, h^i_i=0 $. The equations of motions determined 
by $S_g+S_{em}$ are the coupled Einstein-Maxwell equations of motion:
\begin{align}\label{einsteineq}
\Box h_{\mu\nu} &=-\kappa T^{\textrm{em}}_{\mu\nu}\,, \\
\partial_\mu \left(F^{\mu\nu}-\frac{\alpha^2}{45m_e^4}[4F^2 F^{\mu\nu}+7(F\cdot\tilde{F}) \tilde{F}^{\mu\nu}]\right) &=\kappa\partial_\mu[h^{\mu\beta}F_{\beta}^{\nu}-h^{\nu\beta}F_{\beta}^{\mu}]\,,
\label{maxwelleq}
\end{align}
where we made use of the fact that the electromagnetic field tensor is traceless and defined 
$F^2\equiv F_{\mu\nu} F^{\mu\nu}$ and $\tilde{F} F\equiv\tilde{F}_{\mu\nu} F^{\mu\nu}$, the indices here are raised by flat 
metric tensor $\eta_{\mu\nu}$.

In equation \eqref{maxwelleq} the electromagnetic field tensor, $F_{\mu\nu}$, is the sum of the free field (incident wave) tensor $f_{\mu\nu}$ and the static external field tensor ${F}^{(e)}_{\mu\nu}$, $F_{\mu\nu}={F}^{(e)}_{\mu\nu}+f_{\mu\nu}$, where $|{F}^{(e)}_{\mu\nu}|\gg |f_{\mu\nu}|$. At this point one can see  that the second and the third terms  (the Heisenberg-Euler ones)
in equation \eqref{maxwelleq} modify the usual vacuum Maxwell equations creating 
refraction indexes in external magnetic field, which  give rise to birefringence effects \cite{Brezin:1971nd}.
For the transverse and parallel modes these indexes are equal to
\begin{align} \label{vacuum indexes}
n_1^2-1 &=4\rho B_e^2\sin^2\phi\,,\qquad \textrm{transverse mode}\,, \\ \nonumber
 %\label{n1}
n_2^2-1 &=7\rho B_e^2\sin^2\phi\,,\qquad \textrm{parallel mode}, %\label{n2}
\end{align}
where $B_e$ is the strength of the external magnetic field, $e$ is the electron charge, $\phi$ is the angle between the incident wave and the direction of the external magnetic field $\mathbf{B}_e$, and $\rho$ is defined as,
\begin{equation}
\rho=(\alpha/45\pi)(e/m_e^2)^2.
\label{rho}
\end{equation}
Deviation of the refraction index from unity destroy equality of the photon momentum, $k$, 
and frequency, $\omega$, and gives rise to 
effective photon mass, $m_\gamma^2 = \omega^2 -  k^2 \neq 0$, as one can see from the
solution of the homogeneous part of equation \eqref{maxwelleq}.

Since we are working in the TT gauge the spatial parts of equations \eqref{einsteineq} and \eqref{maxwelleq} 
describing propagating waves now read
\begin{align}
\Box h_{ij} &=-\kappa T^{\textrm{em}}_{ij}  \label{system1}\,,  \\  %\nonumber
\left[\Box- m_\gamma^2 \right]A^j &=\kappa\partial_i[h^{ik}{F}_{k}^{(e)j}-h^{jk}{F}_{k}^{(e)i}]\,,
\label{system2} 
\end{align}
where in the r.h.s. of eq.~(\ref{system1}) we took into account only terms which are bilinear 
in $F_{\mu\nu}$ and $f_{\mu\nu}$, see below eq. (\ref{T-i-j}). 

Let us consider now plane gravitational wave propagating through a region with magnetic field vector $\mathbf{B}_e$
assuming the latter to be homogeneous at the scale of the gravitational wave length. 
We expand as usually the gravitational wave tensor in its Fourier components:
\begin{equation}\label{Furierexph}
 h_{ij}(\mathbf{x}, t)=\sum_{\lambda=\times, +} h_\lambda(\mathbf{x}) \textrm{e}_{ij}^\lambda e^{-i\omega t},
\end{equation}
where $\lambda$ denotes the GW polarization index and $\textrm{e}_{ij}$ is the gravitational wave polarization tensor, which is defined as
\begin{equation}
\textrm{e}_{ij}^+(\hat{\mathbf{x}})=u_iu_j-v_iv_j,\qquad  \textrm{e}_{ij}^\times(\hat{\mathbf{x}})=u_iv_j+v_iu_j,
\end{equation}
where $u, v$ are unit vectors orthogonal to the direction of the 
wave propagation $\hat{\mathbf{x}}$ and to each other. The energy-momentum tensor of the electromagnetic wave 
generated in the process of the graviton-photon transformation is given by:
\begin{equation}
 T_{i}^j=E_iE_{e}^j+B_iB_{e}^j-\frac{1}{2}\delta_{i}^j(E^2+B^2),
\label{T-i-j}
\end{equation}
where lower index $e$ refers to the external electromagnetic field. 

Let us introduce vector potential, $A^j$, of the electromagnetic wave: 
\begin{equation}\label{FurierexpA}
 A_j= i\sum_\lambda\textrm{e}_j^\lambda(\hat{\mathbf{x}}) A_\lambda e^{-i\omega t},
\end{equation}
where 
%an arbitrary phase $i$ was introduced and 
 $\textrm{e}_j^\lambda$ is the photon polarization vector
% the electromagnetic  energy-momentum 
% tensor in the case of a static external magnetic fields assumes the form
%\begin{equation}
% T_{ij}=i\sum_{\lambda=\times, +}\partial_{\mathbf{n}}A_\lambda(\mathbf{n})B_{e\bot}\textrm{e}_{ij}^\lambda e^{-i\omega t},
%\end{equation}
and the magnetic field of the propagating electromagnetic wave is given by $B_k=(\nabla\times A)_k=\epsilon_{ijk}\partial_jA_k$.  
Now plugging  equations \eqref{Furierexph} and \eqref{FurierexpA} into equations \eqref{system1} and \eqref{system2} 
we obtain the following system of equations:
\begin{align}\label{system3}
(\omega^2+\partial_{\mathbf{x}}^2)h_{\lambda} &=-i\kappa \partial_{\mathbf{x}}A_\lambda(\mathbf{x})B_{T}\,,  \\  %\nonumber
(\omega^2+\partial_{\mathbf{x}}^2)A_\lambda+(k^2-\omega^2)A_\lambda &=-i\kappa\partial_{\mathbf{x}}h_\lambda(\mathbf{x})B_{T}\,,
\label{system4}
\end{align}
where $B_T$ is the strength of the transverse external magnetic field. 

The system of equations \eqref{system3}, \eqref{system4} is not easy to handle but 
one can simplify the work assuming that the coherence length of the background magnetic field $\lambda_B$ 
is much greater than the photon wavelength $\lambda_p$: $\lambda_B\gg \lambda_p$. Under this assumption the 
operator $\omega^2+\partial_{\mathbf{x}}^2$ can be expanded as 
$\omega^2+\partial_{\mathbf{x}}^2=(\omega+i\partial_{\mathbf{x}})(\omega-i\partial_{\mathbf{x}})
\simeq 2\omega(\omega+i\partial_{\mathbf{x}})$  where $(-i\partial_{\mathbf{x}}) =k $ is the momentum  operator 
and we assume that refraction index $n$ slightly differs from unity, 
$|n-1|\ll 1$, and $\omega+k\simeq 2\omega$ with $\omega$ satisfying the general dispersion equation $k=\omega n$.   In this case 
the system of equations \eqref{system3}, \eqref{system4} becomes
\begin{align}\label{system5}
(\omega+i\partial_{\mathbf{x}})h_{\lambda} &\simeq-\frac{\kappa}{2} A_\lambda(\mathbf{x})B_{T}\,, \\  %\nonumber
(\omega+i\partial_{\mathbf{x}})A_\lambda+\omega(n-1)A_\lambda & \simeq -\frac{\kappa}{2} h_\lambda(\mathbf{x})B_{T}\,,
\label{system6}
\end{align}
where $n$ is the total refraction index. It includes respectively the QED effects due to vacuum polarization, the plasma 
effects due to refraction of the photon in the medium and birefringence effects such as the Cotton-Mouton effect,
\begin{equation}\label{totalindex}
n=n_\textrm{QED}+n_\textrm{plasma}+n_\textrm{CM}\,.
\end{equation}
Here the plasma refraction index is given by
\begin{equation}
n_\textrm{plasma}=-\frac{\omega_\textrm{plasma}^2}{2\omega^2}\,,
\label{n-plasma}
\end{equation}
where the plasma frequency is as usually $\omega_{plasma}^2 = n_e e^2/m$ with $e^2 = 4\pi\alpha$,
and $n_e$ is the number density of free electrons.

The refraction indices for two polarizations states of photon, $n_1$ and $n_2$, are given by equation \eqref{vacuum indexes} and 
in the case of weak magnetic field they can be approximated  as
\begin{align}\label{vacuum indexes1}
n_+  &=1+\frac{4}{2}\rho B_e^2\sin^2\phi\,,\qquad \textrm{transverse mode\,,}\\\nonumber
n_\times &=1+\frac{7}{2}\rho B_e^2\sin^2\phi\,,\qquad \textrm{parallel mode},
\end{align}
where $n_1=n_+$ and $n_\times=n_2$. The Cotton-Mouton effect arises when the photons travel through gas-like medium and as a consequence  the difference between the two refraction indices is given by
\begin{equation}
n_\textrm{CM}^+-n_\textrm{CM}^\times=C\lambda_p B_e^2,
\end{equation} 
where $C$ is the Cotton-Mouton constant.

The system of equations \eqref{system5}, \eqref{system6} can be written in the matrix form:
\begin{equation}\label{matrix}
\left[(\omega+i\partial_{\mathbf{x}})+
\begin{bmatrix}
  \omega(n-1)_{+} & B_{T}/m_{Pl} & 0 & 0 \\
  B_{T}/m_{Pl}  & 0 & 0 & 0 \\
  0 & 0 & \omega(n-1)_{\times}  &  B_{T}/m_{Pl}\\
  0 & 0 & B_{T}/m_{Pl} & 0
 \end{bmatrix}
\right] 
\begin{bmatrix}
  A_+({\mathbf{x}}) \\
  h_+({\mathbf{x}}) \\
  A_\times ({\mathbf{x}})\\
  h_\times ({\mathbf{x}})
 \end{bmatrix}
=0\,,
\end{equation}
which will be the starting point of the next section.

\section{Graviton-photon mixing} \label{sec:3}

In the previous section we have derived the equations of motion for the graviton-photon system in  
presence of an external magnetic field. In order to solve equation \eqref{matrix} it is necessary to make 
some assumption on the nature of the magnetic field. The system of equations \eqref{matrix} was derived in the approximation of a background magnetic field with coherence length much larger than the photon or graviton wavelength. System \eqref{matrix} can be further simplified by making some reasonable assumptions on the nature of the background magnetic field. In this section we
assume that the background magnetic field is homogeneous on a sufficiently large
coherence length $\lambda_B$. 

In order to solve the system of equations \eqref{matrix} notice that there is no mixing between the photon 
or graviton states $+$ and $\times$. Correspondingly system \eqref{matrix} of four equations decouples into   
two independent systems of two equations each:
\begin{equation}\label{matrix1}
\left[(\omega+i\partial_{\mathbf{x}})+
\begin{bmatrix}
   M_{+} & {0} \\
  {0}  & M_\times
 \end{bmatrix}
\right] 
\begin{bmatrix}
  \hat\Psi_+({\mathbf{x}}) \\
  \hat\Psi_\times({\mathbf{x}}) 
   \end{bmatrix}
=0,
\end{equation}
where we have defined
\begin{equation}
\hat\Psi_+({\mathbf{x}})\equiv 
\begin{bmatrix}
 A_+({\mathbf{x}}) \\
  h_+({\mathbf{x}}) 
  \end{bmatrix} 
 ,\qquad
 \hat\Psi_\times({\mathbf{n}})\equiv 
  \begin{bmatrix}
 A_\times({\mathbf{x}}) \\
  h_\times({\mathbf{x}}) 
 \end{bmatrix} \,,
 \end{equation}
\begin{equation}\label{M-matrix}
M_+\equiv 
\begin{bmatrix}
 m_+ & m_{g\gamma} \\
 m_{g\gamma} & 0  
  \end{bmatrix} \,,
 \qquad
 M_\times\equiv 
\begin{bmatrix}
 m_\times & m_{g\gamma} \\
 m_{g\gamma} & 0 \end{bmatrix} \,,
 \end{equation}
and 
\begin{equation}\label{mixingcomp}
m_+=\omega(n-1)_+, \quad m_\times=\omega(n-1)_\times, \quad m_{g\gamma}=B_T/m_\textrm{Pl}.
\end{equation}

Since there is no mixing between + and $\times$ states, we can
concentrate on one of the reduced matrices,  $M_+$, 
where from now we drop index +. In this case the Schr\"{o}dinger-like equation of motion, to be solved, has the form
\begin{equation}\label{waveq}
(\omega+i\partial_{\mathbf{x}})\hat\Psi(\mathbf{x})+M\hat\Psi(\mathbf{x})=0.
\end{equation} 
Equation \eqref{waveq} can be solved using the unitary transformation of field $\hat\Psi({\mathbf{x}})$:
\begin{equation}
\hat\Psi '({\mathbf{x}})= U\hat\Psi({\mathbf{x}}),
\end{equation}
where $U$ is the unitary matrix with the entries:
\begin{equation}
U=
\begin{bmatrix}
\cos\theta & \sin\theta\\
-\sin\theta & \cos\theta
\end{bmatrix}\,.
\end{equation}
In the new basis the equation of motion reads
\begin{equation}\label{waveq1}
(\omega+i\partial_{\mathbf{x}})\hat\Psi'(\mathbf{x})+M'\hat\Psi'(\mathbf{x})=0,
\end{equation}
where $M'$ is the diagonal matrix, $M'=\textrm{diag}[m_1, m_2]$ and $m_1$ and $m_2$ are the eigenvalues of matrix $M$:
\begin{equation}
m_{1, 2}=\frac{1}{2}[m_+\pm \sqrt{m_+^2+4m_{g\gamma}^2}].
\end{equation} 
The formal solution of equation \eqref{waveq1} is given by
\begin{equation}\label{solution}
\hat\Psi'(\mathbf{x})=\exp\{i\int_\mathbf{0}^{\mathbf{x}'}(\omega+M')\textrm d \mathbf{x}'\}\hat\Psi'(\mathbf{0})\,.
\end{equation}
Now we can go back to the old basis by multiplying the left hand side of equation \eqref{solution} by $U^T$ and obtain
\begin{equation}
\label{solution1}
\hat\Psi(\mathbf{x})=\exp\{i\int_\mathbf{0}^{\mathbf{x}'}U^TM'U\,\textrm d \mathbf{x}'\}\hat\Psi(\mathbf{0})\,,
\end{equation}
where common phase $e^{i\omega |\mathbf x|}$ was absorbed in field $\hat\Psi$. 
The explicit expressions for photon field, $A$, and graviton field, $h$, are
\begin{align}
\label{field-system-A}
A(\mathbf x) &=(\cos^2\theta e^{im_1|\mathbf x|}+\sin^2\theta e^{im_2 |\mathbf x})A(\mathbf 0)+
\sin\theta\cos\theta(e^{im_1|\mathbf x|}-e^{im_2|\mathbf x}|)h(\mathbf 0)\,, \\ %\nonumber
h(\mathbf x) &= \sin\theta\cos\theta(e^{im_1|\mathbf x|}-e^{im_2|\mathbf x|})A(\mathbf 0)+(\sin^2\theta e^{im_1|\mathbf x|}+
\cos^2\theta e^{im_2 |\mathbf x|})h(\mathbf 0)\,,
\label{field-system-h}
\end{align}
where $\theta$ is the graviton-photon mixing angle defined as:
\begin{equation}
\tan 2\theta=\frac{2m_{g\gamma}}{m_+}.
\label{tan-2theta}
\end{equation}

At this point one can easily calculate the probability of the graviton conversion to  photon 
by assuming that initially there are only gravitons and no photons, that is, $h(\mathbf 0)=1$ and $A(\mathbf 0)=0$:
\begin{equation}
\label{oscprobability}
P_{g\rightarrow\gamma}=|\langle h(\mathbf 0)|A(\mathbf x)\rangle|^2=\sin^2(2\theta)\sin^2(\sqrt{m_+^2/4+m_{g\gamma}^2}\cdot |\mathbf x|)\,.
\end{equation}
Equation \eqref{oscprobability} gives the oscillation probability of a graviton to convert into a photon and vice-versa.   We can also 
notice that the expression for the oscillation probability is completely analogous to the oscillation probability between neutrinos with 
different flavors \cite{Dolgov:2002wy}.

\section{Mixing strength: qualitative description}\label{sec:4}

In the previous section we have calculated the probability of graviton to photon transformation and in this section we study 
various regimes of equation \eqref{oscprobability}. For an order of magnitude
estimate, we neglect for the moment the  absorption or 
%scattering terms which arise due to absorption or 
scattering of the photons in the surrounding medium, 
%i.e $-i\Gamma$ where $\Gamma$ is the scattering rate, 
the expansion of the Universe, 
%$iH$ where $H$ is the Hubble constant 
and  the Cotton-Mouton effect. 

In order to estimate the oscillation probability we present the numerical values of the three terms 
in the right hand side of equation \eqref{totalindex}. The plasma effects are included in term 
$m_{\textrm{plasma}}=-\omega_{\textrm{plasma}}^2/2\omega$ and its numerical value 
is\footnote{From now on we omit index + in $m_+$}:
\begin{equation}
m_{\textrm{plasma}}=-3.5\cdot 10^{-17}\left[\frac{1 \textrm{eV}}{\omega}\right]\left[\frac{n_e}{ \textrm{cm}^{-3}}\right]\quad {\textrm{cm}^{-1}},
\label{m-plasma}
\end{equation}
where $n_e$ is the electronic number density. 
The QED effects are included in the term $m_{\textrm{QED}}$\footnote{{In fact the
expression for $m_{\textrm{QED}}$ should include the factor 2 or 7/2 depending on the mode considered in the
calculations. Here we omit these factors since in the case considered below the plasma effects dominate over the
QED effects and the contribution from these factors are not essential}} which reads:
\begin{equation}
m_{\textrm{QED}}=\left[\frac{\alpha}{45\pi}\right]\left[\frac{B_T}{B_c}\right]^2\omega,
\end{equation}
where $B_c=  m_e^2/e = 4.41\cdot 10^{13}$ Gauss
and the numerical value of $m_{\textrm{QED}}$ is:
\begin{equation}
m_{\textrm{QED}}=1.33\cdot 10^{-27}\left[\frac{\omega}{1 \textrm{eV}}\right]\left[\frac{B_T}{1 \textrm{G}}\right]^2\quad {\textrm{cm}^{-1}}.
\end{equation}
The mixing term is $m_{g\gamma}=B_T/m_{Pl}$ and it is equal to:
\begin{equation}
m_{g\gamma}=8\cdot 10^{-26}\left[\frac{B_T}{1 \textrm{G}}\right]\quad \textrm{cm}^{-1}.
\label{m-g-gamma}
\end{equation}
Expression \eqref{oscprobability} has different limiting forms, depending on the value of mixing angle $\theta$.\\
a) \emph{Weak mixing}\\
In this case $\theta\ll 1$ which corresponds to $m_{g\gamma}\ll m$ (remind that $m$ is either $m_+$ or $m_\times$). 
Equation \eqref{oscprobability} in this case becomes
\begin{equation}
\label{weak-mixing}
P_{g\rightarrow\gamma}(|\mathbf x|, \theta)=4\theta^2\sin^2\left[\frac{m |\mathbf x|}{2}\right]=
4\theta^2\sin^2\left[\frac{\pi |\mathbf x|}{l_\textrm{osc}}\right],
%\label{P-g-gamma}
\end{equation}
where the oscillation length is defined as $l_\textrm{osc}=2\pi/m$. Now if the oscillation length is greater than 
path $|\mathbf x|$, the oscillation probability is given by the simple expression:
\begin{equation}
P_{g\rightarrow\gamma}(|\mathbf x|, \theta)\simeq(m_{g\gamma} |\mathbf x|)^2.
\label{P-g-gamma}
\end{equation}
It is interesting to see when the weak mixing condition is fulfilled during the evolution of the Universe. 
In other words we need to check when $m_{g\gamma}\ll m$ that is:
\begin{equation}\label{weakcondition}
8\cdot 10^{-26}\left[\frac{B_T}{1 \textrm{G}}\right]\ll   \mid 1.33\cdot 10^{-27}\left[\frac{\omega}{1 \textrm{eV}}\right
]\left[\frac{B_T}{1 \textrm{G}}\right]^2-
3.5\cdot 10^{-17}\left[\frac{1 \textrm{eV}}{\omega}\right]\left[\frac{n_e}{ \textrm{cm}^{-3}}\right] \mid \,.
\end{equation}
Evidently the l.h.s. of this relation vanishes at the frequency equal to:
\begin{equation}
\omega_{res} = 1.6\cdot 10^5\,{\rm eV}\, \left(\frac{1G}{B_T}\right)\, \left(\frac{n_e}{\rm cm^{-3}} \right)^{1/2}\,.
\label{omega-ras}
\end{equation}
This is the so called resonance frequency when the mixing angle is close to  $\pi/4$. 

Let us see whether the mixing could be weak or strong in the present day universe.
To this end we need to know three parameters which are the strength of magnetic field, $B_T$, the frequency 
of gravitons, $\omega$, and the electronic density, $n_e$. Large scale magnetic fields are constrained by the CMB 
observations since they can create an anisotropic pressure which in turn requires an anisotropic 
gravitational field in order to maintain equilibrium. Gravitational instabilities in the post recombination era, 
 created by the magnetic fields generate fluctuations in the CMB spectrum due to the Sachs-Wolfe effect \cite{Durrer:1999bk}. 
In ref.   \cite{Barrow:1997mj} the authors, using the 4-year Cosmic
Background Explorer (COBE) microwave background
isotropy measurements infer an upper limit on large scales magnetic field strength  $B(t_0)$\footnote{From now we 
omit { index "$T$"} in $B_T$, where the magnetic field strength refers to the transverse part.}
\begin{equation}
B(t_0)\simeq 5\times 10^{-9} f^{1/2} (\Omega_0 h_{70}^2)^{1/2}\quad \textrm{G}\,.
\label{B-max}
\end{equation}
where $f$ is a shape factor of the order of unity. Recent limits based on the WMAP 7 year and South Pole 
Telescope (SPT) data, allow to conclude that
the primordial magnetic field on scales, $\lambda_B < 1$ Mpc  is bounded by $B(t_0)\lesssim 3$ nG at 95\% (CL) \cite{Paoletti:2012bb}. 

Primordial magnetic field also induces the Faraday rotation of the linear polarization of 
CMB and can induce non zero parity odd cross correlations between the CMB temperature 
and B-polarization anisotropies. The authors of ref. \cite{Kahniashvili:2008hx} put upper limits on 
the amplitude of the large scale magnetic field 
in the range $6\cdot 10^{-8}$ G to $2\cdot 10^{-6}$ G. 
More stringent constraints on large scale magnetic fields at the present day wave length $\lambda_B\sim 0.1$ Mpc 
come from BBN bound on gravitational waves. If primordial magnetic field was generated before BBN, it 
would create an anisotropic stress in the l.h.s. of the
Einstein equations which in turn would create perturbations in the curvature of space-time, namely GWs. 
The authors of ref. \cite{Caprini:2001nb} argued that the large scale magnetic fields, produced  at 
the electroweak phase transition must be weaker than $B\lesssim 10^{-27}$ G and 
the magnetic field produced at inflation weaker than$B\lesssim 10^{-39}$ G. 
These results were criticized in ref. \cite{Kosowsky:2004zh}. 
In this paper we assume validity of the CMB bounds on the large scale magnetic  fields, quoted above.

At the present epoch the free electron number
density is not a well known quantity and just for an order of magnitude
estimate we take it equal to its upper bound, assuming that almost all matter is ionized
\begin{equation}\label{redshift}
n_e(t_0)\lesssim n_B(t_0) =\frac{3H_0^2\Omega_B}{8\pi Gm_p}=1.123\cdot 10^{-5}(h_0^2\Omega_B)\,\textrm{cm}^{-3}\,\simeq 2.47\cdot 10^{-7}\,\textrm{cm}^{-3}\,,
\end{equation}
where according to WMAP 7 years measurements \cite{Komatsu:2010fb} $H_0=100h_0$ km/s/Mpc 
with $h_0\simeq 0.7$; $h_0^2\Omega_B\simeq 0.022$ is the present day baryon density parameter and $m_p$ is the proton mass.
More accurate estimates are presented below.

According to eq. \eqref{weakcondition}, the validity of the weak mixing condition depends upon the photon frequency. 
As a guiding example let us take  $\omega=10^3$ eV, $B(t_0)\simeq 5\cdot 10^{-9}$ G and $n_e(t_0)\simeq 2.47\cdot 10^{-7}$ \textrm{cm}$^{-3}$ and obtain:
\begin{equation}
l_\textrm{osc}=\frac{2\pi}{|m_{\textrm{plasma}}|}\simeq 7.26 \cdot 10^{26} \textrm{cm},
\end{equation}
where clearly the plasma effects dominates over QED effects and $m\simeq |m_{\textrm{plasma}}|$. 
So for the path of the graviton of the order of the present day Hubble radius $|\mathbf x|=H^{-1}(t_0)\simeq 1.32\cdot 10^{28}$ cm,  
the oscillation probability today would be:
\begin{equation}\label{weakprob}
P_{g\rightarrow\gamma}\simeq 2\cdot 10^{-15}.
%\label{P-today}
\end{equation}
Here we used eqs. (\ref{tan-2theta}, \ref{m-plasma}, \ref{m-g-gamma}, \ref{weak-mixing}).
Equation \eqref{weakprob} shows that for the present day value of the magnetic field $B(t_0)$ equal to its upper bound \eqref{B-max}, 
$n_e(t_0)$ determined by eq. \eqref{redshift}, and for the gravitons with frequencies $\omega\ll m_e$ the condition 
for the weak mixing regime is satisfied.  
Thus the probability of graviton transition to photon is small but not negligible, which can lead to some observable effects. 

It is  worth noting 
that for photons with frequencies $\omega\gg m_e$ the Euler-Heisenberg Lagrangian is no longer applicable for the calculations of
the photon refraction index in external electromagnetic field.  In the limit of high energies factor $\rho$ (\ref{rho}) is a function of the graviton energy. This would lead to resonance transition after recombination and to larger probability of photon production.
The case when $\omega\gg m_e$ will be considered elsewhere.

It is instructive to estimate the graviton-photon transition probability at different periods of  the cosmological evolution. 
Before the matter-radiation decoupling at $z\simeq1090$, one might expect that the magnetic field was larger 
than that at the recombination because under condition of the magnetic flux conservation the field strength evolves
as the inverse scale factor squared.  This would be true if the cosmological
magnetic field was generated at some earlier epoch, before recombination.

If magnetic field was generated before the BBN era, its strength could be constrained by 
the observed abundances of light elements. In particular,  an impact of magnetic field on the cosmological expansion,
an increase of  the decay rate of neutrons, and other  phenomena, described e.g. in
ref. \cite{Giovannini:2003yn}, would change the abundances of light elements.
In the pioneering papers  \cite{Matese:1969zz} the  upper limit on the field strength 
 at BBN was derived: $B\lesssim10^{12}$ G.
 In more recent studies of the effect of magnetic field on the abundances of light 
 elements, especially of $^4$He,  somewhat weaker bound,  $B\lesssim10^{13}$ G, was inferred~\cite{Kernan:1996ab}. 
So huge magnetic fields are formally allowed at BBN. 

On the other hand, the electronic number density increases with decreasing scale factor as $n_e(a)\sim a^{-3}\sim T^3$. 
This leads to an increase of the plasma effects, so they dominate over the QED effects. Thus in this case the weak mixing 
condition is realized,  as one can see from equation \eqref{weakcondition}. The very small mixing angle  gives  negligible 
transition probability  of gravitons to photons  with frequencies $\omega\ll m_e$.

Things start to change near recombination when the plasma temperature was $T\simeq 0.26$ eV. The electronic density
(plasma density) can be parametrized as
\begin{equation}
n_e(t_\textrm{rec})=X_e \, n_B(t_0)(T/T_0)^3,
\label{ne-rec}
\end{equation}
where $n_B$ is the total baryon density, $n_B=n_p+n_H$, with $n_p$ and $n_H$ being respectively the free proton and neutral 
hydrogen densities, $X_e (z)$ is the red-shift dependent ionization fraction, defined as $X_e=n_p/n_B$. The condition of
electric neutrality, $n_e = n_p = X_e n_B$, is of course assumed.
Since the present day baryon number density is given  by equation \eqref{redshift}, we obtain
\begin{equation} \label{redshift1}
n_e(t_\textrm{rec})\simeq 1.123\cdot 10^{-5} X_e(1+z)^3(h_0^2\Omega_B)\, \textrm{cm}^{-3}\,,
\end{equation}
where $T/T_0=1+z$ is substituted. Ionization fraction, $X_e(z)$, can be calculated by  solving the
out of equilibrium Saha-like non linear differential equation. Near recombination time, the solution is  well approximated 
by the expression   \cite{Jones:1985}:
\begin{equation}\label{rec_ionization}
X_e(z)=7.2\cdot 10^{-3}\frac{(h_0^2\Omega_M)^{1/2}}{h_0^2\Omega_B}\left(\frac{1+z}{1090}\right)^{12.75},\quad (800<z<1200),
\end{equation}
where $h_0^2\Omega_M$ is the present day matter density parameter. Inserting equation \eqref{rec_ionization} into 
equation \eqref{redshift1} we get
\begin{equation}
n_e(t_\textrm{rec})\simeq 104.71\,(\Omega_M h_0^2)^{1/2}\left(\frac{1+z}{1090}\right)^{15.75}\,\textrm{cm}^{-3}\quad (800<z<1200).
\end{equation}
Since according to WMAP 7 year data
the redshift at recombination is $1+z=1090$ and the matter density parameter is $\Omega_M h_0^2\simeq 0.15$, the free electron density
at recombination would be $n_e(t_\textrm{rec})\simeq 40.5\,\textrm{cm}^{-3}$.

Assuming that the observed contemporary magnetic field originated from primordial magnetic field seeds without much dynamo effects, 
one finds that on the horizon length at the recombination time, the field strength was:
\begin{equation}\label{B-rec}
B(t_\textrm{rec})\simeq B(t_0)(1+z)^{2}=3\cdot 10^{-3}\, \textrm{G},
\end{equation}
where we took $B(t_0)\simeq 3\cdot 10^{-9}$. 

Taking $B(t_{\textrm{rec}})\sim 3\cdot 10^{-3}$~G and $n_e\sim 40.5$ cm$^{-3}$, 
we find  that the ratio
of the plasma term to the QED term at recombination:
\begin{equation}
r=\frac{|m_{\textrm{plasma}}|}{m_{\textrm{QED}}}=2.63\cdot 10^{10}\left[\frac{1 \textrm{eV}}{\omega}\right]^2
\left[\frac{1 \textrm{G}}{B}\right]^2\left[\frac{n_e}{ \textrm{cm}^{-3}}\right]
\label{r}
\end{equation} 
is larger than unity for all frequencies below
$\omega\lesssim 3.4\cdot 10^8$ eV. Recall that the approximation used here is valid 
only for $\omega \ll m_e $.

With these values we estimate the oscillation probability at the recombination
epoch when the Hubble radius was equal to: 
\begin{equation}\label{hubbleconst}
H(t_\textrm{rec})^{-1}=H_0^{-1}/[\Omega_\Lambda+\Omega_M(1+z)^3]^{1/2}\simeq 6.7\cdot 10^{23} \textrm{cm},
%\label{H-rec}
\end{equation}
where $H_0^{-1}=1.32\cdot 10^{28}$ cm, $\Omega_\Lambda\simeq 0.7$ and $\Omega_M\simeq 0.3$. The oscillation length in this case reads
\begin{equation}
l_\textrm{osc}=\frac{2\pi}{|m_{\textrm{plasma}}|}\sim 4.43 \cdot 10^{20}\, \textrm{cm},
\end{equation}
which is 3 orders of magnitude smaller than the Hubble distance at recombination.
Taking $B(t_\textrm{rec})\simeq 3\cdot 10^{-3}$ G, $n_e(t_\textrm{rec})\simeq 40.5$ cm$^{-3}$,
and graviton initial energy $\omega_i\simeq 10^5$ eV 
and using eqs.~(\ref{tan-2theta}, \ref{m-plasma}, \ref{m-g-gamma}, \ref{weak-mixing}),
we find that the oscillation probability is
\begin{equation}\label{probrec}
P_{g\rightarrow\gamma}\simeq 10^{-15} \,\left(\frac{\omega_i}{10^5\,{\rm eV}}\right)^2
\,\left(\frac{B_i}{3\cdot 10^{-3}\,{\rm G}}\right)^2\,
\,\left(\frac{40.5\, {\rm cm}^{-3}}{n_e}\right)^2.
\end{equation}
One can see that  probability \eqref{probrec} depends on the frequency of the graviton and  
noticeable amount of high energy photons can be produced if the original graviton spectrum is not cut-off at high frequencies \cite{Dolgov:2011cq}.\\ 
b) \emph{Resonance}\\
At the resonance $r=1$, eq. (\ref{r}), and thus $m=0$, 
so the mixing angle becomes large, 
$\theta = \pi/4 $. In this case the expression for the oscillation probability is
\begin{equation}
P_{g\rightarrow\gamma}=\sin^2(m_{g\gamma}\cdot |\mathbf x|).
\end{equation}
If the resonance is wide the complete transition of  graviton into  photon is possible.
Note that near the resonance the two regimes of weak mixing and maximum mixing have the same expression for 
probability (\ref{P-g-gamma})
in the case when the oscillation length is larger than the path. 
However, the excitation of resonance depends upon the effects of damping and loss of coherence and so its proper treatment
demands the density matrix formalism.  This is done in the next section. 

Even if the resonance is not excited, as is the case when $\omega \ll m_e$, the density matrix formalism 
leads to an essential enhancement  of  the photon production because the photons loose coherence due to
scattering on electrons and do not oscillate back. This happens if the coherence loss rate
is faster than the Universe expansion rate.  

\section{Oscillations: density matrix description}\label{sec:5}

In the previous section we have calculated the probability of the graviton-photon oscillations in the wave 
function approximation. Graviton conversion into photons in the present day universe in the wave function approximation was also considered in ref. \cite{Pshirkov:2009sf} but their results are significantly different from ours as we will see below. This approximation is sufficiently accurate if the loss of coherence due to non-forward
or inelastic scattering of the participating particles (i.e. of the photons in the considered case) may be neglected.
This is realized if the mean free path with respect to such scattering is greater than the oscillation length. In the
opposite case the graviton-photon system becomes open (i.e. not self-contained) and the density matrix formalism 
should be applied. The corrections to the wave function approximation are especially important in the resonance 
situation when the oscillation length becomes large:
\begin{equation}
l_{osc} = \frac{2\pi}{\sqrt{ m^2/4 + m_{g\gamma}^2 }} \rightarrow \frac{2\pi}{\sqrt{m_{g\gamma}^2 }}\,,
\label{losc-res}
\end{equation}
where  $m = 0$ at resonance.

Generally speaking the density matrix operator the is $4\times 4$-matrix describing transitions between photon and gravitons
with different helicity states:
\begin{equation}
\hat\rho\equiv |A_+, h_+, A_\times, h_\times\rangle\otimes\langle A_+, h_+, A_\times, h_\times|,\end{equation}
where the C-valued density matrix $\hat\rho_{ij}$ is obtained by averaging the matrix elements over medium, 
$\rho=\langle\hat\rho\rangle$.  However, since there is no mixing between states $|+\rangle$ and $|\times\rangle$ 
the density matrix is reduced to two independent $2\times 2$-matrices separately for  $|+\rangle$ and $|\times\rangle$ 
states having the form
\begin{equation}
\hat\rho=\hat\Psi\hat\Psi^\dagger,
\end{equation}
where $\hat\Psi$ is two-dimensional column describing the graviton and photon states $\Psi = [\Psi_g,\Psi_\gamma]^T$ of 
either polarization $|+\rangle$ or $|\times\rangle$. Here upper index "$T$" means transposition.

The density matrix  operator satisfies the Liouville-von Neumann equation: 
\begin{equation}
i\frac{d \hat\rho}{ d t}=[\hat{\mathcal H}, \hat\rho]
\end{equation}
where $\mathcal H$ is the total Hamiltonian of the system. If the system is open the total Hamiltonian is not Hermitian 
and the anti-Hermitian part of the Hamiltonian describes coherence breaking due to scattering and absorption. 
In general case it is expressed through the collision integral modified to include the matrix structure of the
process. For the case of neutrino oscillations the equation for the density matrix was derived in ref.~\cite{ad-nu-osc},
see also~\cite{rho-nu}. It is a non-linear  integro-differential equation due to presence of complicated collision integrals.
However, for an an order of magnitude estimate the equation can be linearized in the usual way\footnote{As usually  $[,]$ denotes the commutator between two operators $a$ and $b$ and $\{,\}$ denotes the anti-commutator. }:
\begin{equation}\label{Liouvileeq}
i\frac{d \rho}{ d x}=
%\hat{\mathcal H}\hat\rho-\hat\rho\hat{\mathcal H}^\dagger=
[ {M}, \rho]-i\{\Gamma, \left(\rho - \rho_{ext}\right)\},
\end{equation}
where $ M$ is given by eq. \eqref{M-matrix}, $\rho_{ext}$ is the density matrix of the corresponding particles in the
medium, the time derivative has been replaced with derivative respect to position in space, $d/dt=d/dx$ for $c=1$ and $\Gamma$ is the damping factor which in our case has the form:
\begin{equation}
\Gamma =
\begin{bmatrix}
\Gamma_\gamma & 0\\
0 & 0
\end{bmatrix}
\end{equation}
where $\Gamma_\gamma$ is the inverse mean free path due to Thompson scattering of photons on 
electrons $\Gamma_\gamma=\sigma_T n_e$ with $\sigma_T=6.65 \cdot 10^{-25}$ cm$^{-2}$ being the Thompson cross section. The damping of
gravitons is neglected due to weakness of their interactions. We also neglect $\rho_{ext}$, assuming that the medium is not populated 
by photons. The latter can be easily taken into account.

In the previous section we estimated the oscillation probability at various stages during the evolution of the Universe, 
namely at the present time and at the recombination but the universe expansion was not explicitly accounted for.
To do that in the cosmological  FRW metric we notice that for a given function $f(p, t)$ the total derivative is given by:
\begin{equation}
\frac{df }{dt} = \frac{\partial f}{\partial t} + \dot p \frac{\partial f}{\partial p} = 
 \frac{\partial f}{\partial t} - H p \frac{\partial f}{\partial p} \,.
 \label{df-dt}
 \end{equation}
Here we  have taken into account the redshift, $\dot p = -Hp$, where  $H= \dot a/a$ is the Hubble parameter
and $a$ is the cosmological scale factor.

Hence eq. (\ref{Liouvileeq}) can be rewritten as
\begin{equation}
iHa\frac{\partial\rho}{\partial a}=[ {M}, \rho]-i\{\Gamma, \rho\},
\end{equation}
where we made use of the fact that in the FRW metric $\partial_x=Ha\partial_a$.

Let us write the off-diagonal density matrix elements  as:
$\rho_{\gamma g}=\rho_{g\gamma}^*=R+iI$, where $R$ is the real part and $I$ is the 
imaginary part. The diagonal components $\rho_{gg}$ and $\rho_{\gamma\gamma}$ are the number densities of gravitons and photons, 
respectively, so they are real and non-negative.
Thus after the split between the real and imaginary parts we obtain the following system of differential equations:
\begin{eqnarray}\label{densitysys}
\rho_{\gamma\gamma}' &=&\frac{-2m_{g\gamma}I - \Gamma_\gamma\, \rho_{\gamma\gamma}}{Ha},\\
\rho_{gg}' &=& \frac{2m_{g\gamma}I}{Ha} ,\\
R'&=& \frac{mI-\Gamma_\gamma R/2}{Ha}\label{R1} ,\\
I'&=& \frac{-mR-\Gamma_\gamma I /2 - m_{g\gamma}(\rho_{gg}-\rho_{\gamma\gamma})}{Ha}
\label{densitysys1},
\end{eqnarray}
where prime means derivative with  respect to $a$ and the
initial conditions are taken at the initial value of the scale factor $a=a_i$ as:
$\rho_g(a_i)=1, \rho_{\gamma}(a_i)=0, I(a_i)=0$, and $R(a_i)=0$.
To solve this system of equations we need to know how  parameters 
$m, m_{\gamma g}$, and $\Gamma_\gamma$ (in units $\textrm{cm}^{-1}$) depend on the scale factor:
\begin{eqnarray}
m_{\gamma g}(a) &=& 8 \cdot 10^{-26}\left[\frac{B_i}{1\textrm{G}}\right]\left[\frac{a_i}{a}\right]^2,\\
\Gamma_\gamma(a)&=& 2.12 \cdot 10^{-22}X_e(a)\left[\frac{a_i}{a}\right]^3,\\
m(a)&=&1.33 \cdot 10^{-27}\left(\frac{B_i}{1 \textrm{G}}\right)^2\left(\frac{\omega_i}{1\textrm{eV}}\right)\left(\frac{a_i}{a}\right)^5-1.12 \cdot 10^{-14}X_e(a)\left(\frac{1\textrm{eV}}{\omega_i}\right)\left(\frac{a_i}{a}\right)^2,
\end{eqnarray}
where initial values $B_i$ and $\omega_i$ are taken at the cosmological recombination time, $t_i=t_{\textrm{rec}}$ and we expressed $n_e(a)=X_e(a)n_B(t_{\textrm{rec}})$ with $n_B(t_{\textrm{rec}})=n_B(t_0)(1+z)^3\simeq 320 \textrm{ cm}^{-3}$. 
{The ionization fraction is governed by the following differential equation  \cite{Weinberg:2008}:
\begin{equation}\label{eqioniz}
\frac{\mathrm d X_e}{\mathrm d a}=-\frac{\alpha n_B}{Ha}\left(1+\frac{\beta}{\Gamma_{2s}+8\pi/\lambda_{\alpha}^3n_B(1-X_e)}\right)^{-1}\left(\frac{SX_e^2+X_e-1}{S}\right),
\end{equation}
where $H$ is the Hubble parameter, $\Gamma_{2s}=8.22458$ s$^{-1}$ is the two-photon decay rate of $2s$ hydrogen state, 
$\lambda_{\alpha}=1215.682\cdot 10^{-8}$ cm is the wavelength of Lyman $\alpha$ photons, $\alpha(T)$ is the case B recombination 
coefficient and $S(T)$ is the coefficient in the Saha equation, $X(1+SX)=1$. Both $\alpha$ and $S$ depend on temperature $T$ 
which can be expressed in terms of scale factor $a$ as follows:
\begin{equation}
T=\left(\frac{g_{*S}(T_i)}{g_{*S}(T)}\right)^{1/3}\left(\frac{a_i}{a}\right)T_i
\end{equation}
where $g_{*S}(T)$ is number of the entropy degrees of freedom. Generally it depends on temperature but
after recombination the only effective massless particles in the standard model are 3 neutrino species and the photon. 
Thus the number of the degrees of freedom is constant which is equal to $g_{*S}(T)=3.91$. 
In this case the temperature drops down as $T=T_i/a$ where $T_i=(1+z_{\textrm{rec}})T_{0\gamma}=2970.25$ K. 
In what follows we  take $a_i =1$ corresponding to recombination time, $t_i=t_{\textrm{rec}}$.
Coefficient $\alpha$ depends on the scale factor as \cite{Hummer1994}:
\begin{equation}
\alpha(a)=\frac{1.038\cdot 10^{-12}a^{0.6166}}{1+0.352a^{-0.53}},
\end{equation}
while  $S(a)$ is equal to
\begin{equation}
S(a)=6.221\cdot 10^{-19}e^{53.158a}a^{-3/2}.
\end{equation}
Coefficient $\beta$ which is also a function of temperature can be expressed through $\alpha$ as follows
\begin{equation}
\beta(a)=3.9\cdot 10^{20}a^{-3/2}e^{-13.289a}\alpha.
\end{equation}

With the these parameters equation \eqref{eqioniz} was solved numerically. Some values of the ionization fraction 
$X_e(a)$ as a function of the scale factor are presented in \autoref{tab:1} and plots of $X_e(a) $ and $X_e(T)$
are shown in \autoref{fig:Fig. 1}. Below we solve numerically equations describing evolution of the mixed graviton-photon density
matrix together with equation (\ref{eqioniz}) introducing into them a factor describing reionization of the cosmic plasma. 
\begin{table}[h]
\begin{center}
\begin{tabular}{| p{4cm} | p{4cm} |}
\hline
$a$ & $X_e(a)$\\  \hline                        
  1 & $0.13$  \\
  1.04 & $0.0813$  \\
  1.18 & $0.0160$\\
  1.23 & $0.00947$\\
  1.5 &$0.00171$\\
  2 &$0.000731$\\
  4 & $0.000373$\\
  10 & $0.000278$\\
  30 & $0.000246$\\
  51.9 & $0.000239$\\
  136.25 & $1$\\
  1090& 1\\
  \hline  
\end{tabular}
\caption{Ionization fraction $X_e$ as a function of the scale factor for $h_0^2\Omega_M\simeq 0.15$ and 
$h_0^2\Omega_B\simeq 0.022$ calculated from eq.~\eqref{eqioniz}. We have also presented two additional pieces
of data for $a=136.3$, which correspond to the period of reionization, and to the present day $a=1090$ where the 
intergalactic medium is almost fully ionized with constant ionization fraction, $X_e\simeq 1$}. 
\label{tab:1}
\end{center}
\end{table}

\begin{figure}[h]

\begin{subfigure}[b]{0.3\textwidth}
  \centering
  \includegraphics[scale=0.8]{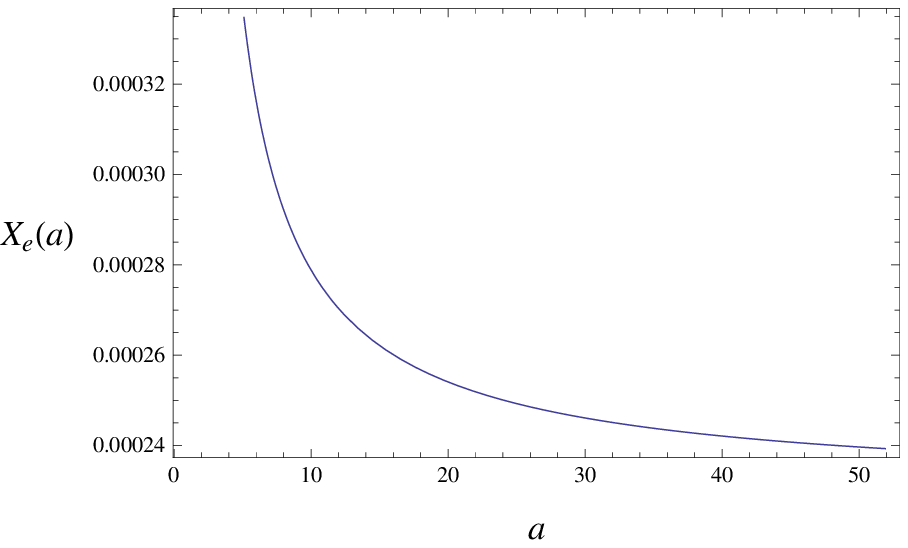}
  \caption{}
   \label{fig:Fig1}
    \end{subfigure}\hspace{3.1cm}
     \begin{subfigure}[b]{0.3\textwidth}
  \centering
   \includegraphics[scale=0.8]{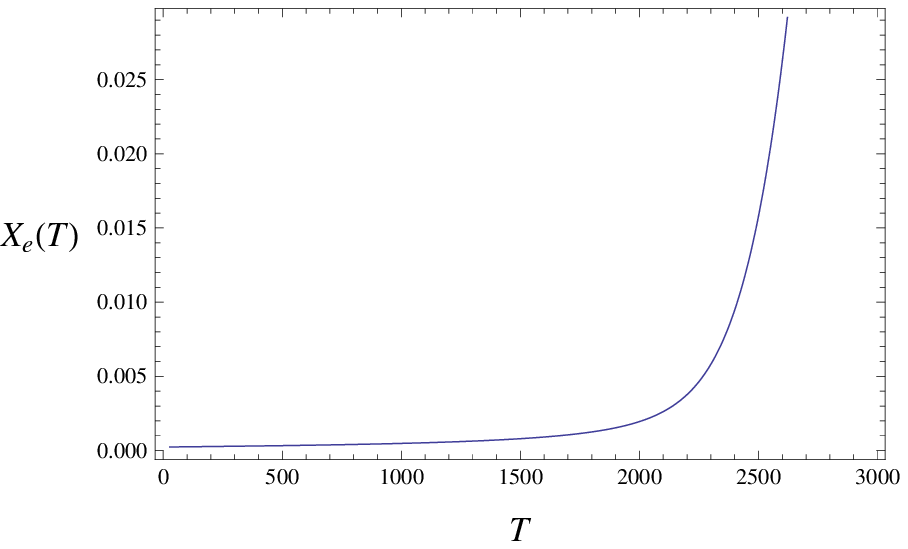}
    \caption{}
    \label{fig:Fig2}
        \end{subfigure}%
        ~ %add desired spacing between images, e. g. ~, \quad, \qquad etc. 
          %(or a blank line to force the subfigure onto a new line)
 
        ~ %add desired spacing between images, e. g. ~, \quad, \qquad etc. 
          %(or a blank line to force the subfigure onto a new line)
   \caption{Ionization fraction $X_e$ as a function of cosmological scale factor $a$ (\ref{fig:Fig1}) and as a function of 
   temperature $T$ in Kelvin (\ref{fig:Fig2}). In panel
   (\ref{fig:Fig1}) the plot is shown in the interval starting from recombination, $a=1$, till $a=51.9$ corresponding to the
   beginning of reionization. The same thing is shown in panel (\ref{fig:Fig2}), where temperature $T$ changes in the interval 
   $T=2970.25$ K corresponding to recombination time to $T=57.22$ K corresponding to onset of of reionization.}
   \label{fig:Fig. 1}
\end{figure}

The resonance condition, $m(a)=0$, is fulfilled at  
\begin{equation}
\omega = 2.9\,{\rm MeV}\, X_e^{1/2}(a)\left(\frac{1{\rm G}}{B_i}\right)\, a^{3/2}\,.
\label{omega-res}
\end{equation}
The ratio of the oscillation length to the mean free path in this case is  
\begin{equation}
\frac{l_{osc}^{(res)}}{l_{free}} = \frac{2\pi\cdot\Gamma_\gamma}{m_{g\gamma}} = 1.66\cdot 10^4 \left[\frac{1\,\rm G}{B_i}\right].  
\label{losc-lfree}
\end{equation}
In this case the resonance frequency at recombination would 
be about 10 MeV for $B_i\simeq 10^{-1}$ G. For possibly larger $B_i$ resonance shifts to smaller $\omega$. 
Even out of the resonance inequality $l_{osc} \geq l_{free}$ may remain true and use of the density matrix formalism is 
obligatory.  

We need to mention however, that very energetic gravitational waves with energy $\omega\gg m_e$ would 
create photons whose scattering on electrons
is weaker than the Thompson one, roughly speaking,  by  $(m_e/E)^2$. This effect would diminish the damping factor 
$\Gamma_\gamma$ by the same amount and can be easily taken into account. One has however to keep in mind that
the Heisenberg-Euler approximation is valid only for $\omega \ll m_e$ and thus we should not go beyond that value.

In the process of the cosmological expansion the graviton-photon transition could pass through resonance
if their frequency satisfies resonance condition (\ref{omega-res}). However, since we consider graviton energies, $\omega\ll m_e$, the resonance condition \eqref{omega-res} is not satisfied in the post-recombination epoch.

Next we need  to evaluate  factor $Ha$. Since, we are interested in the cosmological epoch just after recombination till the present time, when the Universe is dominated by nonrelativistic matter, the Hubble constant as a function of the redshift is given by equation \eqref{hubbleconst}. 
{ In terms of the scale factor, the product $Ha$ is given by
\begin{equation}
Ha=H(t_i)\,[\Omega_M/a+\Omega_\Lambda a^2]^{1/2}.
\end{equation}
For redshift $z>1$ we can neglect the contribution of the cosmological constant into the energy density of the 
Universe since it becomes important only for $z\lesssim 1$. We take
$\Omega_M\simeq 1$ and $1/Ha = 6.7\cdot 10^{23} a^{1/2} $ cm for the former case 
and  $1/Ha = 6.7\cdot 10^{23} [0.3/a+0.7a^2]^{-1/2} $ cm for the latter case. According to our 
notations the redshift $z=1$ corresponds to the scale 
factor $a=545$ with respect to the recombination time.

For $7< z<11$ the universe is re-ionized by the first generation  stars. 
According to ref. \cite{Dunkley:2008ie} if the universe went into a sudden complete ionization, the re-ionization redshift  $z_{\textrm{reion}}=6$ is excluded at 99\% (CL) in favor of $z_{\textrm{reion}}=11$. Using the WMAP 5 year 
data the  authors of ref. \cite{Dunkley:2008ie} suggest that the universe underwent an extended period of partial re-ionization 
starting at $z\sim 20$ and ending with a complete ionization for $z\sim 7$ instead of a sudden re-ionization. 
For $z\sim 20$ the value of the scale factor with respect to recombination is $a=51.9$ and for $z=7$, $a=136.25$.
 
In Fig. \ref{fig:Fig. 2}, Fig. \ref{fig:Fig. 3}, and Fig. \ref{fig:Fig. 4} the probability of photon creation, $\rho_{\gamma\gamma}$ is
presented as a function of the cosmological scale factor for various values of the initial background magnetic field and for graviton energy $\omega_i=10^5$ eV. For such $\omega$, 
the resonance does not occur but still the probability is much higher than the simple estimate in the wave function formalism. 
We can see that in all Fig. \ref{fig:Fig. 2}, Fig. \ref{fig:Fig. 3}, and Fig. \ref{fig:Fig. 4} the oscillation probability rapidly increases for $a<20$ and 
remains almost constant for $a>20$ until onset of the period of reionization at $a\simeq 52$. 
The rapid increase of 
$\rho_{\gamma\gamma}$ for $a<20$ is due to two reasons: a quick drop of the
ionization fraction from its value at recombination and sharp decrease of the oscillation frequency. 
For $20<a<52$ the ionization fraction practically remains constant with $\rho_{\gamma\gamma}$ slowly rising. From beginning of 
the reionization period at $a\simeq 52$ till $a\simeq 545$ or $z\simeq 1$, 
$\rho_{\gamma\gamma}$ slowly decreases with superimposed oscillations of decreasing amplitude. 
For $a>545$ the vacuum energy density dominates over the matter energy density. At this period
the photon creation probability remains almost constant. The oscillation probability drops from $a\simeq 100$ up to $a=1090$ roughly speaking by 20-30\,\%. The value of the magnetic field at recombination has been evaluated by the anti-redshift of the present day large scale magnetic field as given in ref. \cite{Paoletti:2012bb, Kahniashvili:2008hx}.}
\begin{figure}[!t]
 \begin{subfigure}[b]{0.3\textwidth}
  \centering
   \includegraphics[scale=0.8]{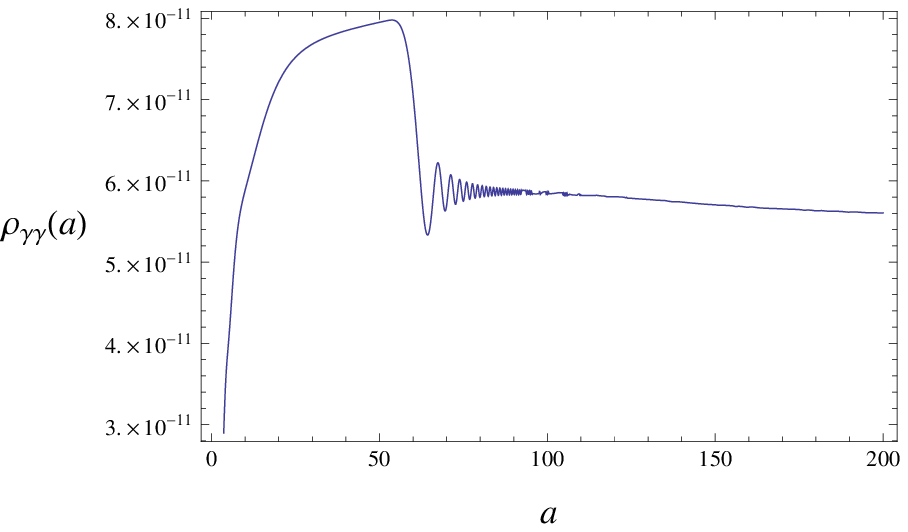}
    \caption{}
    \label{fig:Fig3}
        \end{subfigure}\hspace{3.1cm}%
        ~ %add desired spacing between images, e. g. ~, \quad, \qquad etc. 
          %(or a blank line to force the subfigure onto a new line)
 \begin{subfigure}[b]{0.3\textwidth}
  \centering
  \includegraphics[scale=0.8]{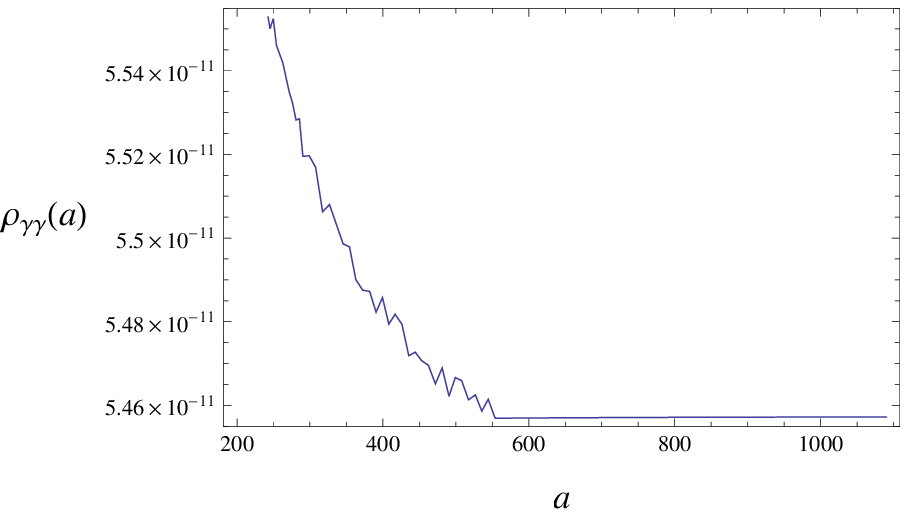}
  \caption{}
   \label{fig:Fig4}
    \end{subfigure}        ~ %add desired spacing between images, e. g. ~, \quad, \qquad etc. 
          %(or a blank line to force the subfigure onto a new line)
   \caption{Probability $P_{g\gamma}=\rho_{\gamma\gamma}$ of photon production by graviton as a function of cosmological scale factor 
   $a$ for graviton initial energy $\omega_i=10^{5}$ eV and background magnetic filed $B_i\simeq 3\cdot 10^{-3}$ G . 
   In (\ref{fig:Fig3}) it is shown the probability in the interval for $a$ from recombination until $a=200$ and in (\ref{fig:Fig4}) 
   it is shown for $a>200$, where it remans almost constant.  }
   \label{fig:Fig. 2}
\end{figure}

\begin{figure}[!t]
 \begin{subfigure}[b]{0.3\textwidth}
  \centering
   \includegraphics[scale=0.8]{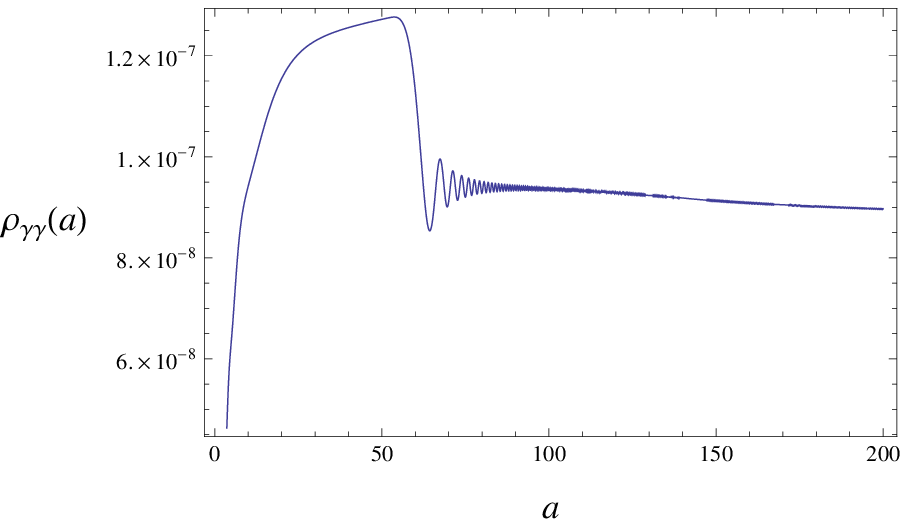}
    \caption{}
    \label{fig:Fig5}
        \end{subfigure}\hspace{3.1cm}%
        ~ %add desired spacing between images, e. g. ~, \quad, \qquad etc. 
          %(or a blank line to force the subfigure onto a new line)
 \begin{subfigure}[b]{0.3\textwidth}
  \centering
  \includegraphics[scale=0.8]{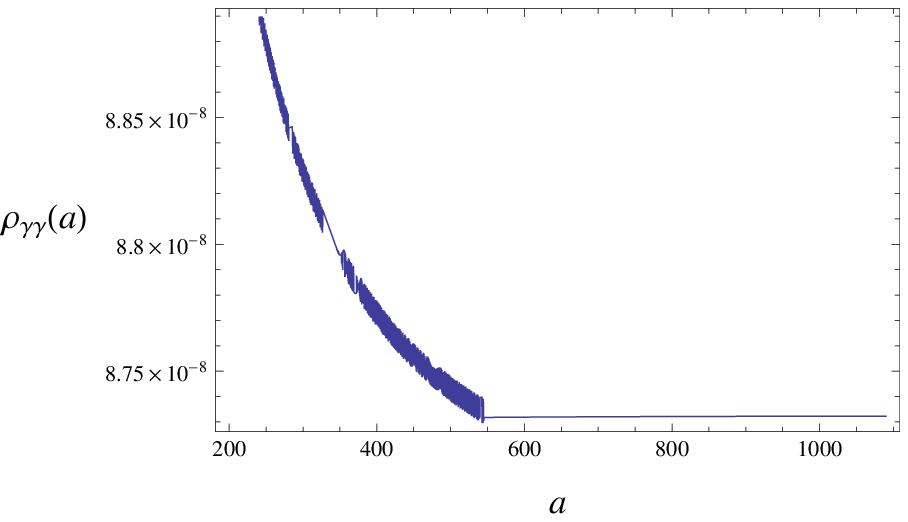}
  \caption{}
   \label{fig:Fig6}
    \end{subfigure}        ~ %add desired spacing between images, e. g. ~, \quad, \qquad etc. 
          %(or a blank line to force the subfigure onto a new line)
   \caption{Probability $P_{g\gamma}=\rho_{\gamma\gamma}$ of photon production by graviton as a function of cosmological scale factor $a$ for graviton initial energy $\omega_i=10^5$ eV and for background magnetic field $B_i\simeq 1.2\cdot 10^{-1}$ G. 
   The plotting interval has been split as in \autoref{fig:Fig. 2}.}
   \label{fig:Fig. 3}
\end{figure}
\begin{figure}[!t]
 \begin{subfigure}[b]{0.3\textwidth}
  \centering
   \includegraphics[scale=0.8]{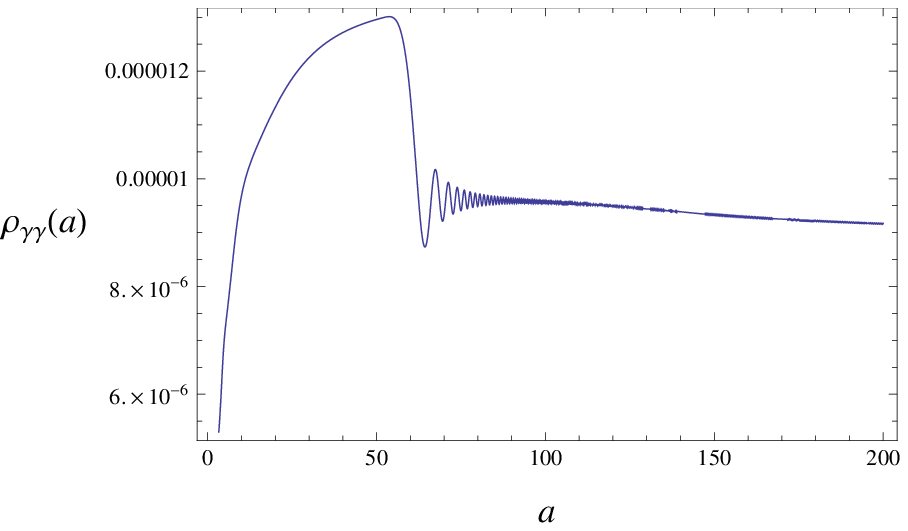}
    \caption{}
    \label{fig:Fig7}
        \end{subfigure}\hspace{3.1cm}%
        ~ %add desired spacing between images, e. g. ~, \quad, \qquad etc. 
          %(or a blank line to force the subfigure onto a new line)
 \begin{subfigure}[b]{0.3\textwidth}
  \centering
  \includegraphics[scale=0.8]{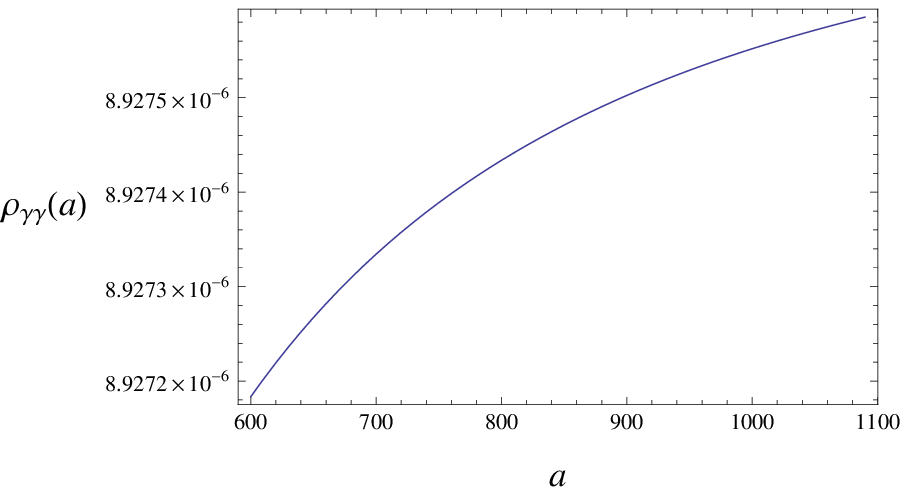}
  \caption{}
   \label{fig:Fig8}
    \end{subfigure}        ~ %add desired spacing between images, e. g. ~, \quad, \qquad etc. 
          %(or a blank line to force the subfigure onto a new line)
   \caption{Probability $P_{g\gamma}=\rho_{\gamma\gamma}$ of photon production by graviton as a function of cosmological scale factor $a$ for initial magnetic field $B_i\simeq 1.2$ G and graviton energy $\omega_i=10^5$ eV.  In (\ref{fig:Fig7}) is shown the oscillation probability for $a\lesssim 200$ and in (\ref{fig:Fig8}) it is shown only for $a>545$ or $z<1$. In the latter we can clearly see an almost constant $\rho_{\gamma\gamma}$ with a slowly variation with the scale factor. }
   \label{fig:Fig. 4}
\end{figure}

\section{Models of an early production of high frequency gravitons. }\label{sec:6}

In the previous section we calculated the probability of graviton-to-photon transition 
taking into account both redshift and coherence breaking in plasma and found that for graviton energy of the 
order of $\omega\simeq 0.1$ MeV the oscillation probability is quite large, $P_{g\gamma}\simeq 10^{-11}$ for $B\simeq 10^{-3}$ G up to $10^{-5}$ for $B\simeq 1.2$ G.  
The number density of the produced photons, which could be directly observed as X-ray background, is
proportional to the initial density of the gravitons. 
The amount of GWs at present time is usually expressed through the density parameter in gravitational waves which is defined as
\begin{equation}
\Omega_{\textrm{gw}}(f, t_0)=\frac{1}{\rho_c}\frac{\mathrm d\rho(f, t_0)}{\mathrm d(\log f)},
\end{equation}
where $\rho_c$ is the present day critical energy density:
\begin{equation}
\rho_c=3m_{\textrm{Pl}}^2H_0^2/8\pi=1.878\cdot10^{-29}h^2\,\, {\rm g/cm}^{3}\,,
\label{rho-c}
\end{equation}
$f=\omega/2\pi$ is the present day GW frequency and $t_0$ is the universe age. 
Since we are interested in GWs of cosmological origin, we consider here only those emitted before BBN. 

The abundances of light elements produced at BBN depend upon the energy density of relativistic species
at $t \sim 1-100$ sec, see e.g. book~\cite{Kolb:1990vq}.  According to the recent data~\cite{bbn-extra-nu} 
an additional energy density at BBN, equal to that of one massless neutrino is allowed and even desirable,
$\Delta N_\nu \leq 1$. The particles which carry this additional energy are not known. They are called generically dark 
radiation. At BBN the energy density of one neutrino species (that is of neutrino plus antineutrino with vanishing 
chemical potential) is approximately equal to that of photons. However after $e^+e^-$-annihilation the ratio of neutrino
to photons energy densities dropped  down approximately by factor four. Keeping in mind that the contemporary
energy density of CMB photons is $h_0^2\Omega_{\textrm{CMB}} \approx 2.5\cdot 10^{-5}$, we find that the total energy density
of the gravitational waves integrated over their spectrum in the present day universe cannot exceed 
\begin{equation} 
h_0^2 \Omega_{GW}(t_0) \lesssim 5.61\cdot10^{-6} \Delta N_\nu\,,
\label{BBN-GW-limit}
\end{equation}
where $\Delta N_\nu $ is the allowed by BBN number of the effective neutrino species. Equation (\ref{BBN-GW-limit})
is an absolute bound on the energy density of all GWs produced before BBN. It would be interesting if $\Delta N_\nu \approx 1$
is explained by primordial GWs.

The oscillation probability strongly depends on the graviton frequency and spectrum.
The models of primordial GW production mostly predict low frequency of stochastic background of GWs, mainly concentrated at 
the present day frequencies of GWs near $f\lesssim1$ Hz. 
For example, inflationary models predict an almost scale invariant spectrum  at large wavelengths, and their density parameter depends mainly on two factors, the GW frequency and the Hubble parameter $H$ at inflation. In the frequency range $f=10^{-15}$ Hz up to $f\simeq$ GHz the density parameter is very low, $h_0^{2}\Omega_{\textrm{gw}}(t_0)\sim 10^{-15}$. Other post inflationary models such as pre-heating 
phase~\cite{Easther:2006gt}, first order phase transitions~\cite{Grojean:2006bp}, and topological defects~\cite{Mazumdar:2010pn}, 
in particular, cosmic strings~\cite{Hogan:2006we} predict in the high frequency range $f\sim$ GHz the 
density parameter of the order $h_0^{2}\Omega_{\textrm{gw}}(t_0)\lesssim 10^{-8}$.

All the above mentioned GWs production models, though predict a substantial density parameter, have maximum frequency today 
not more than $10^{-5}$ eV.  We calculated numerically the graviton-photon oscillation probability for frequencies 
$f\sim 10^{-5}$ eV and found that it is of  the order, $10^{-30}$. With this low value of the  probability the total 
density parameter in photons of most of post-inflationary GWs models at the maximum frequency $f\sim$ GHz would be
\begin{equation}
h_0^{2}\Omega_{\gamma}(t_0)=h_0^{2}\Omega_{\textrm{gw}}(t_0)\cdot\rho_{\gamma\gamma}\simeq 10^{-38}.
\end{equation}
Such a small value of the density parameter makes improbable observations of photons from these GWs.

However there is a particular model of GWs emission during the cosmological time interval between the 
Big Bang and the BBN epoch which leads to rather high cosmological energy density of the very energetic GWs.
It was suggested in ref. \cite{Dolgov:2011cq, Anantua:2008am} that after Big Bang the Universe could have passed 
a transient stage of matter domination by very light primordial black holes of mass $M<10^8$ g, which would   
completely evaporate before BBN leaving no trace. The number density of such light BHs is not constrained by any 
observational data and during their domination it could reach value of the order of unity. 
In particular, in ref. \cite{Dolgov:2011cq} 
different mechanisms of GWs emission are considered, where the produced amount of GW could exceed that produced by 
other mechanisms. Among the mechanisms considered there, we single out the graviton evaporation, 
where the emitted peak frequency of quasi-thermal gravitons would be in the range from $f\sim$ keV up to $f\sim$ MeV today. 
The peak frequency in this model depends on the BH mass which turns out to be in the interval from the Planck mass, 
 up to $\lesssim10^{8}$ g.

Let us consider, for example, gravitons with the initial energy 
$\omega_i=\omega(t_{\textrm{rec}})=10^5$ eV. Their frequency today would be 
$\omega=\omega_i/(1+z_{\textrm{rec}})=\omega_i/1090=91.7$ eV and they would produce 
photons with the same frequency.
The value of the graviton density parameter at this frequency according to the above quoted scenario, could be
$h_0^{2}\Omega_{\textrm{gw}}(t_0)\simeq5\cdot 10^{-8}$. And so
the corresponding density parameter of the produced  photons could be in the interval(or even two orders of magnitude higher):
\begin{equation}
 h_0^{2}\Omega_{\gamma}(t_0)=h_0^{2}\Omega_{\textrm{gw}}(t_0)\cdot P_{g\gamma}\simeq 3\cdot 10^{-18}-4\cdot 10^{-13},
\end{equation}
where $P_{g\gamma}\simeq 5.5\cdot 10^{-11}-9\cdot 10^{-6}$ is the probability of graviton to photon conversion. 
%For higher initial energies, $\omega_i\simeq 10^8$ which correspond to $\omega\simeq$ 10 keV at present, 
%the density parameter in gravitons is $h_0^{2}\Omega_{\textrm{gw}}(t_0)\simeq 10^{-8}$ and 
%$P_{g\gamma}\simeq10^{-9}$ would give a density parameter in photons $h_0^{2}\Omega_{\gamma}(t_0)\simeq 10^{-17}$. 
%We can notice that the photon density parameter slowly changes with increased graviton initial energy. This due to 
%the fact that in the energy range, $10^5$ eV up to $10^8$ eV the oscillation probability slowly changes with energy where the resonance 
%width is very narrow. 
The energy flux of such photon background at the present time would be
\begin{equation}
F_{\gamma}=\left(\frac{\textrm d E_{\textrm{gw}}}{\textrm dA\,\mathrm dt}\right)\cdot P_{g\gamma}=
c\,T_{00}^{\textrm{gw}}\cdot P_{g\gamma}=c\,h_0^2\Omega_{\textrm{gw}}\,\rho_c\cdot P_{g\gamma},
\end{equation}
where $T_{00}^{\textrm{gw}}=\rho_{\textrm{gw}}$ is the 00 component of the GW energy-momentum tensor 
and we restored the light velocity in order to express the photon flux in the standard units (erg/cm$^2$\,s). Taking the present day
photon energy $\omega_{\textrm{ph}}\sim 0.1$ keV, $c=3\cdot 10^{10}$ cm/s and 
$\rho_c=1.878\cdot 10^{-29}\,$ gr/cm$^{-3}$ we obtain the energy flux 
\begin{equation}
F_{\gamma}\simeq 1.5\cdot 10^{-15}-2\cdot 10^{-10}\, [\textrm{erg/cm$^2$\,s}],
\end{equation}
which is comparable to the energy flux of most AGNs in the soft X-ray spectrum \cite{Lehman:2001} and even higher.

\section{Discussion and conclusion}\label{sec:7}

We have shown that the probability of the graviton-to-photon transition in large scale cosmological magnetic field after recombination epoch could be in the range $P_{g\gamma}\sim 10^{-10}-10^{-5}$ at frequencies in 0.1 keV range (in the present day values). An efficient oscillations between
graviton and photon could exist at higher frequencies too, but the Heisenberg-Euler approximation, which we use in this work, 
becomes invalid at energies exceeding the electron mass. This will be considered elsewhere.

For smaller frequencies, e.g.  1 eV, the transition probability would be smaller, by about 2-3 orders of magnitude. The oscillation 
probability strongly depends on the external magnetic field and the graviton energy. 
For higher values of these parameters, as our numerical calculations show, the oscillation probability can 
increase by several orders of magnitude. 

The photons produced by such mechanism could make considerable contribution to cosmic electromagnetic background
if the density of the original gravitational waves, $h_0^2\Omega_{\textrm{gw}}$, is sufficiently high.
We have estimated efficiency of the photon production in various models of primordial GW generation 
discussed in the literature. Mostly, in the considered inflationary and post-inflationary models the density of
photons produced by the GWs is quite low and is not observable at the present time.

The mechanism discussed here gives a large number of photons (for fixed values of the magnetic field) only for high frequency gravitons. 
We think that given the present GWs production models the only mechanism before BBN that could generate a 
measurable flux of photons is the graviton production by primordial black holes. 
Since PBH emit thermal gravitons (if one neglects gray body corrections)
the spectrum of the GWs today would be rather close to the original one with some distortion induced by
the different cosmological moments of GW creation, as shown in the first paper of ref.~\cite{Dolgov:2011cq}.

After hydrogen recombination the plasma density drops down  
and the interaction of the photons, created by the graviton-photon transition,
with electrons becomes much weaker in complete analogy with the CMB photons. Such photons could make
observable contributions to the cosmological electromagnetic background, in particular, to X-rays or extragalactic 
light.  If we consider a conservative present day density parameter $h_0^2\Omega_{\textrm{gw}}(t_0)\simeq 10^{-7}-10^{-8}$, the energy flux in X-rays by the proposed mechanism would be of the order $F_\gamma\simeq 10^{-10}-10^{-15}$ erg/cm$^2$/s where the flux upper limit is 10 percent less than the observed energy flux in the soft X-rays \cite{Ajello:2008xb}. If we assume that the total density of the gravitational waves reaches its upper bound allowed by BBN (and explains
the possibly observed dark radiation),
 $h_0^2\Omega_{\textrm{gw}}(t_0)\lesssim10^{-5}$, and that their frequency is close to 0.1 keV, the energy flux today would be in the range,  $F_\gamma\simeq 10^{-8}-10^{-13}$ erg/cm$^2$/s. This energy flux could explain the cosmic X-ray background being its dominant part without requiring any obscured AGN.\\

{\bf Acknowledgements} A. Dolgov  acknowledges the support of the Russian Federation Government Grant 
No. 11.G34.31.0047. D. Ejlli thanks D. Semikoz for useful discussions on the present limits of the large scale magnetic 
fields and the spectrum of extragalactic radiation.

  \end{document}